\documentclass[preprint]{aastex}

\slugcomment{NAOJ-Th-Ap 2001,No.2; ApJL, in press}

\shorttitle{Yutaka Fujita}
\shortauthors{Heating by Blast Waves}

\begin{document}

\title{Heating of X-Ray Hot Gas in Groups by Blast Waves}

\author{Yutaka Fujita}
\affil{National
Astronomical Observatory, Osawa 2-21-1, Mitaka, Tokyo 181-8588, Japan}
\email{yfujita@th.nao.ac.jp}

\begin{abstract}
In order to find the conditions which determine whether X-Ray hot gas in
galaxy groups (intragroup gas; IGG) is heated externally or internally,
we investigate the evolution of blast waves in galaxy groups growing on
a hierarchical clustering scenario. We find that the blast waves driven
by quasars are confined in groups and heat the IGG internally at
$z\lesssim 1$. However, at $z\gtrsim 1$, they expel the IGG from groups;
the expelled gas may fall back into the groups later as externally
heated gas. Moreover, this may explain the observed low metal abundance
of IGG. For blast waves driven by strong starbursts, the shift of the
fate of blast waves occurs at $z\sim 3$. On the other hand, although
blast waves driven by weak starbursts do not expel IGG from groups, the
heating efficiency decreases at $z\gtrsim 3$ because of radiative
cooling. It will be useful to compare these results with XMM-Newton
observations.
\end{abstract}


\section{Introduction}

X-ray properties of clusters and groups of galaxies show the thermal
history of the X-ray gas \citep{kaiser91,evrad91,fujita00}. Simple
theoretical models predict that the relation between X-ray luminosity
and temperature should be $L_X\propto T_X^{2}$, if the thermal
properties of X-ray gas have been determined only by the gravitational
energy released at the time of the collapse. However, X-ray observations
show that this is not true; from a rich cluster scale to a group scale,
the exponent increases from $ L_X\propto T_X^{2-3}$
\citep[e.g.][]{david93,xue00} to $L_X\propto T_X^{5}$
\citep{ponman96,xue00}. Moreover, the discovery of the entropy excess in
groups (`entropy-floor') by \citet{pon99} suggests that
non-gravitational heating has especially affected the thermal properties
of X-ray gas in groups (intragroup medium; IGG).

However, the heating sources have not been identified; they may be
quasars or starburst galaxies \citep[e.g.][]{val99}. In order to know
what are the dominant sources, it may be useful to investigate whether
the entropy excess is the residual of the entropy originally present in
the protocollapse medium or intergalactic medium (IGM) or whether it is
generated within halos after collapse. \citet{toz00} showed that using
XMM-Newton it would be possible to find whether IGG is {\em externally}
or {\em internally} heated by observing entropy profiles at large radii
in X-ray halos. If IGG is externally heated, we will detect the
isentropic, low surface brightness emissions extending to radii larger
than the virial ones in groups. However, even if we detect them, we need
theoretical models to be compared. That is, we need the models
describing what kind of heating source heats IGG externally.

Moreover, the epoch when the energy is released into IGG is still open
to question. \citet{yam00} consider the heating by AGN jets. Using a
simple theoretical model, they estimated the Sunyaev-Zeldovich effect by
the heated gas and compared it with the observations of the Cosmic
Microwave Background. They concluded that the IGG is heated at
$z\lesssim 3$. This suggests that the heating of IGG occurred after or
simultaneously with the collapse of groups. However, they did not
consider the heating by starburst galaxies. Moreover, they assumed that
the energy is ejected into IGM with the density near to the average in
the universe. In the actual universe, it is likely that AGNs reside in
the region with IGM density higher than the average in the universe.

In this paper, for various heating sources we investigate when the
sources heat IGG externally or internally. For that purpose, we consider
the evolution of blast waves with different energies, because it is
expected that heating sources (quasars or starburst galaxies) drive
blast waves similar to supernova remnants \citep{voi96,yam99}. We assume
that the heating sources responsible for the excess entropy in groups
have resided in the groups or in the group progenitors; this assumption
is valid unless each source can heat a extremely wide region of the
universe. We explore whether the blast waves are confined in the groups
(or in their progenitors) or whether they escape from the groups (or
 from their progenitors) and expel the IGG. In the former case, they
serve as internal heating sources; since the hot gas region inside the
wave is buoyant, it later mixes with the ambient gas through
Rayleigh-Taylor instability. Moreover, subsequent mergers between groups
may also make their IGG uniform. Thus, in that case we expect that the
energy released by heating sources is effectively transferred into
IGG. On the other hand, in the latter case, the expelled gas may fall
back into the groups later as externally heated gas as the groups gather
the ambient medium. We also investigate the radiative cooling of the
blast waves. If it is effective, most of the energy ejected by heating
sources is radiated before transferred into IGG.

\section{Models}
\subsection{The Growth of Galaxy Groups}

The conditional probability that a particle which resides in a object
(`halo') of mass $M_2$ at time $t_2$ is contained in a smaller halo of
mass $M_1\sim M_1+d M_1$ at time $t_1$ ($t_1<t_2$) is
\begin{equation}
 \label{eq:prob}
P_{1}(M_1,t_1|M_2,t_2)d M_1
=\frac{1}{\sqrt{2\pi}}
\frac{\delta_{c1}-\delta_{c2}}{(\sigma_{1}^2-\sigma_{2}^2)^{3/2}}
\left|\frac{d\sigma_1^2}{d M_1}\right|
\exp\left[\frac{(\delta_{c1}-\delta_{c2})^2}
{2(\sigma_{1}^2-\sigma_{2}^2)}\right]d M_1 \;,
\end{equation}
where $\delta_{ci}$ is the critical density threshold for a spherical
perturbation to collapse by the time $t_i$, and $\sigma_i [\equiv
\sigma(M_i)]$ is the rms density fluctuation smoothed over a region of
mass $M_i$ for $i=1$ and 2 \citep{bon91,bow91,lac93}. 

We define the typical mass of halos at $t$ that become part of a larger
halo of mass $M_0$ at later time $t_0 (>t)$ as
\begin{equation}
 \label{eq:m_ave}
 \bar{M}(t|M_0,t_0)
=\frac{\int_{M_{\rm min}}^{M_0} M P_{1}(M,t|M_0,t_0)d M}
{\int_{M_{\rm min}}^{M_0} P_{1}(M,t|M_0,t_0)d M} \:,
\end{equation}
where $M_{\rm min}$ is the lower cutoff mass. We choose $M_{\rm
min}=10^8\: M_{\sun}$, which corresponds to the mass of dwarf galaxies.
In the following sections, we investigate the group whose virial mass is
given by
\begin{equation}
 \label{eq:m_vir}
M_{\rm vir}(t|M_0,t_0)=\bar{M}(t|M_0,t_0) \:.
\end{equation}
 From now on, we will represent $M_{\rm vir}(t|M_0,t_0)$ with $M_{\rm
vir}$ unless it is misunderstood.

We assume that groups are spherically symmetric. The virial radius of a
group with virial mass $M_{\rm vir}$ is defined as
\begin{equation}
 \label{eq:r_vir}
r_{\rm vir}=\left(\frac{3\: M_{\rm vir}}
{4\pi \Delta_c(z) \rho_{\rm crit}(z)}\right)^{1/3}\:,
\end{equation}
where $\rho_{\rm crit}(z)$ is the critical density of the universe and
$\Delta_c(z)$ is the ratio of the average density of the group to the
critical density at redshift $z$. The former is given by
\begin{equation}
\label{eq:rho_crit}
 \rho_{\rm crit}(z)
=\frac{\rho_{\rm crit,0}\Omega_0 (1+z)^3}{\Omega(z)}\:,
\end{equation} 
where $\rho_{\rm crit,0}$ is the critical density at $z=0$, and 
$\Omega(z)$ is the cosmological density parameter. The latter
is given by
\begin{equation}
\label{eq:Dc_lam}
  \Delta_c(z)=18\:\pi^2+82 x-39 x^2\:, 
\end{equation}
for the flat universe with cosmological constant \citep{bry98}. In
equation (\ref{eq:Dc_lam}), the parameter $x$ is given by
$x=\Omega(z)-1$. The virial temperature of a group is given by
\begin{equation}
 \frac{k_{\rm B}T_{\rm vir}}{\mu m_{\rm H}}
=\frac{1}{2}\frac{GM_{\rm vir}}{r_{\rm vir}}\:,
\end{equation}
where $k_{\rm B}$ is the Boltzmann constant, $\mu (=0.6)$ is the mean
molecular weight, $m_{\rm H}$ is the hydrogen mass, and $G$ is the
gravitational constant. We assume that IGM had not been affected by
non-gravitational heating until blast waves were driven. Thus, since the
average mass density of a group is given by $\Delta_c \rho_{\rm crit}$,
the average density of the IGG is given by $\rho_{\rm IGG}=f_{\rm
gas}\Delta_c \rho_{\rm crit}$, where $f_{\rm gas}$ is the gas or baryon
fraction of the universe. We use $f_{\rm gas}=0.25 (h/0.5)^{-3/2}$,
where the present value of the Hubble constant is written as
$H_0=100\:h\rm\: km\:s^{-1}\: Mpc^{-1}$. The value of $f_{\rm gas}$ is
the observed gas mass fraction of high-temperature clusters
\citep{moh99,ett99,arn99}, for which the effect of non-gravitational
heating is expected to be small.

\subsection{The Evolution of Blast Waves}

The Sedov-Taylor solution for pointlike explosions adequately describes
the early phase of the evolution of blast waves. It gives a shock radius
of
\begin{equation}
 r_s = \xi\left(\frac{E_0}{\rho_{\rm IGG}}\right)^{1/5}t^{2/5},
\end{equation}
where $\xi=1.15$, $E_0$ is the explosion energy, and $t$ is the time
elapsed since the explosion \citep{spi78}.

If $E_0$ is relatively small, the hot gas region surrounded by a blast
wave becomes in pressure equilibrium with the ambient gas before the
wave escapes from the group. The radius at which the pressure
equilibrium attained is approximately written as
\begin{equation}
 r_p = \left(\frac{3E_0}{4\pi P_a}\right)^{1/3},
\end{equation}
where $P_a$ is the pressure of the ambient gas. We call this radius `the
pressure equilibrium radius'. We assume that the pressure of the ambient
gas is given by
\begin{equation}
P_a=\frac{\rho_{\rm IGG}k_{\rm B}T_{\rm vir}}{\mu m_{\rm H}}.
\end{equation}
On the other hand, if $E_0$ is large or $r_p>r_{\rm vir}$, the blast
wave escapes from the group and the IGG of the group is expelled.

If the density of IGG, $\rho_{\rm IGG}$, is large, radiative cooling may
affect the evolution of blast waves. The postshock temperature is given
by
\begin{equation}
 T_s = \left(\frac{\mu m_{\rm H}}{k_{\rm B}}\right)
\frac{8}{25}\frac{(\gamma-1)}{(\gamma+1)^2}\xi^2
\left(\frac{E_0}{\rho_{\rm IGG}}\right)^{2/5}t^{-6/5},
\end{equation}
where $\gamma=5/3$ is the adiabatic index \citep{spi78}. The postshock
cooling time is given by
\begin{equation}
 t_c=\frac{3}{2}\frac{P_s}{n_e n_i \Lambda(T_s)},
\end{equation}
where $P_s$, $n_e$, and $n_i$, respectively, are the pressure, electron
density, and ion density of the postshock gas, and $\Lambda$ is the
cooling function. We adopt the cooling function of 1/100 solar metal
abundance derived by \citet{sut93}. The cooling becomes important when
$t_c<t_{\rm exp}$, where $t_{\rm exp}=r_s/(dr_s/dt)$ is the expansion
time scale. We define the cooling radius $r_c$ as the one at which the
condition $t_c=t_{\rm exp}$ is satisfied. If $r_c<r_p$, we expect that
most of the energy released by an explosion is radiated and is not
transferred into IGG.

\section{Results}
\label{sec:result}

We adopt a CDM model with $\Omega_0=0.3$, $\Lambda=0.7$, $h=0.7$, and
$\sigma_8=1.0$. Figures~1a-c show the evolutions of $r_p$, $r_s$, and
$r_{\rm vir}$ of a galaxy group with {\em present} mass of
$M_0=10^{14}\;M_{\sun}$.  The input energies are $E_0=10^{61}$,
$10^{59}$, and $10^{56}$~erg, respectively.  We call the model of
$E_0=10^{61}$~erg a `quasar model', because the typical energy of a
quasar activity is $\sim 10^{61}$~erg \citep[e.g.][]{yam99}. Moreover,
we refer to the models of $E_0=10^{59}$ and $10^{56}$~erg as a 'strong
starburst model' and a `weak starburst model', respectively. Note that
the energies correspond to the binding energies of galaxies with the
mass of $1.4\times 10^{11}\; M_{\sun}$ and $1.2\times 10^{9}\;
M_{\sun}$, respectively \citep{sai79}.

In the quasar model, the radii have a relation of $r_p<r_{\rm vir}<r_c$
for $z\lesssim 1$ (Figure~\ref{fig:r}a). Thus, the blast wave is
confined in the group and the explosion energy is effectively
transformed into the IGG. Thus, the IGG is internally heated. On the
other hand, for $z\gtrsim 1$, the radii have a relation of $r_{\rm
vir}<r_c<r_p$, which means that the blast wave escapes from the group,
and the IGG is expelled.

The fate of the blast wave in the strong starburst model is
qualitatively the same as that in the quasar model; the wave is confined
for $z\lesssim 3$ but gets out of the group for $z\gtrsim 3$
(Figure~\ref{fig:r}b). On the other hand, in the weak starburst model,
radiative cooling becomes important for $z\gtrsim 3$ because $r_c<r_p$
(Figure~\ref{fig:r}c). Thus, the heating is inefficient for $z\gtrsim
3$, although it is efficient for $z\lesssim 3$ ($r_c>r_p$).

\section{Discussion}

We have investigated the evolution of blast waves in galaxy groups
growing on a hierarchical clustering scenario. We found that the blast
waves driven by quasars are confined in groups and heat the intragroup
gas (IGG) internally at $z\lesssim 1$. However, at $z\gtrsim 1$, they
expel the IGG; the expelled gas may fall back into the groups later as
externally heated gas. For the blast waves driven by strong starbursts,
the shift of the fate of blast waves occurs at $z\sim 3$. On the other
hand, the blast waves driven by weak starbursts do not expel IGG, and
the heating efficiency decreases at $z\gtrsim 3$ because of radiative
cooling. The results can be used to determine the heating sources of IGG
and the heat input epoch by comparing them with the predictions of
\citet{toz00}.

\begin{figure}
\figurenum{1}
\epsscale{0.60}
\plotone{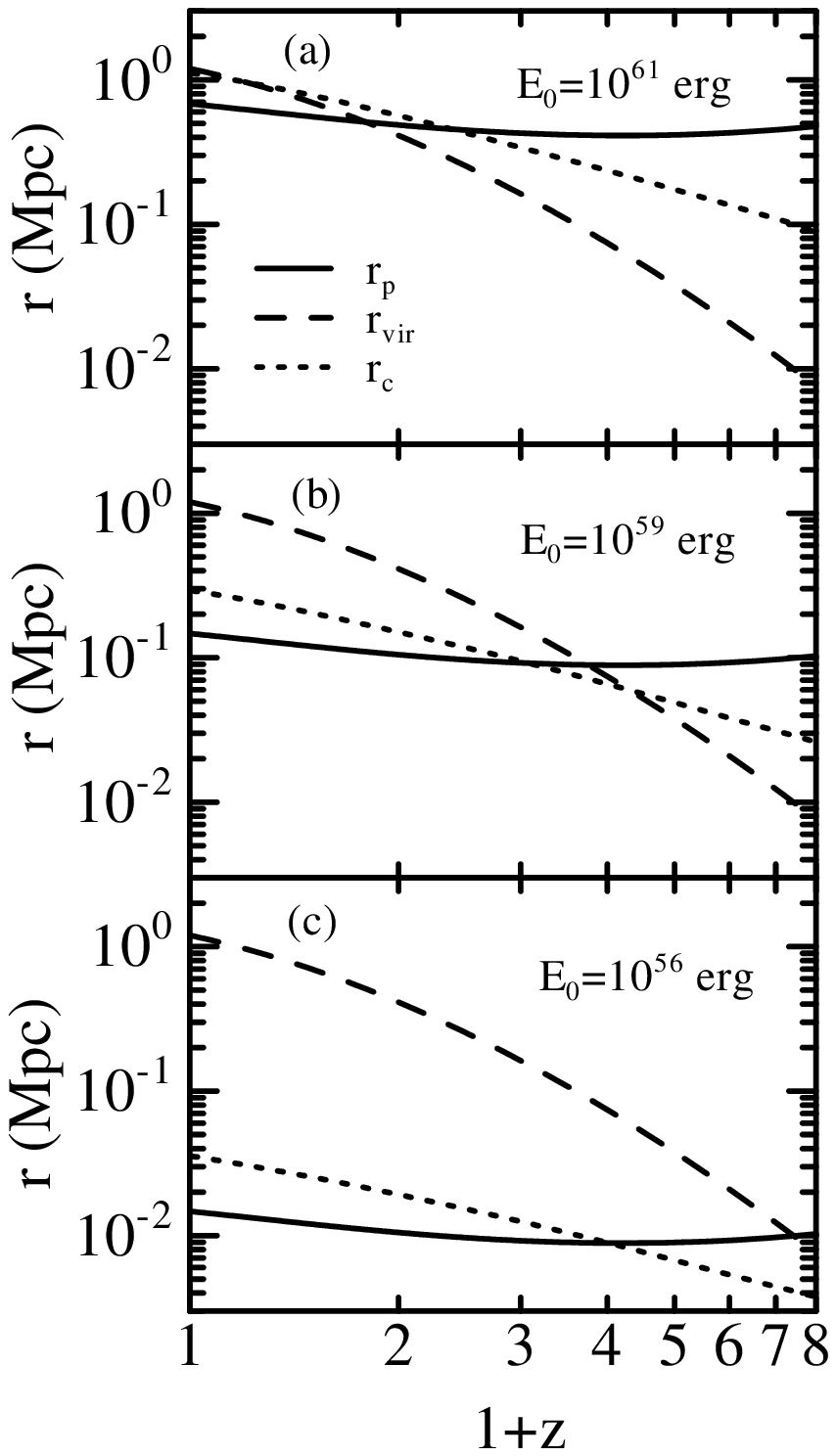}
\caption{The evolutions of the pressure equilibrium radius
(solid lines), virial radius (dashed lines), and cooling radius (dotted
lines) for a group with $M_0=10^{14} M_{\sun}$. (a) $E_0=10^{61}$, (b)
$E_0=10^{59}$, and (c) $E_0=10^{56}$~erg. \label{fig:r}}
\end{figure}

Note that several studies have suggested that the energy input by
supernovae (including starburst galaxies) falls short of the observed
energy injection. Using a simple theoretical model, \citet{val99}
indicated that the energy provided by supernovae cannot raise the
entropy of IGG up to the level required by current
observations. Moreover, \citet{kra00} estimated the energy provided by
supernovae from the observed metal abundance of X-ray gas and found that
the heating only by supernovae requires unrealistically high
efficiency. Thus, quasars or AGNs may be the main contributor of the
heating of IGG \citep{wu00}. If this is the case, the energy input epoch
is expected to be $z\sim 2$ at which the number density of quasars
reaches the maximum \citep{har90,war94,sch95,ken95}. Our quasar model
shows that IGG is expelled from groups by blast waves at $z\sim 2$
(Figure~\ref{fig:r}a). Thus, if quasars have mainly heated IGG, we may
detect the isentropic, low surface brightness emissions extending to
radii larger than the virial ones in groups according to \citet{toz00}.

The IGG expelled by quasars may mix with a large amount of the
intergalactic medium (IGM) surrounding the groups. Thus, if metal is
ejected from galaxies into the IGG at $z\gtrsim 2$, it may be diluted
further with the surrounding IGM. X-ray observations show that the metal
abundance of IGG is small in comparison with that of X-ray gas in
clusters \citep{ren97,fuk97}. The observed IGG with low metal abundance
may be the IGM later accreted by the groups.

\begin{figure}
\figurenum{2}
\epsscale{0.60}
\plotone{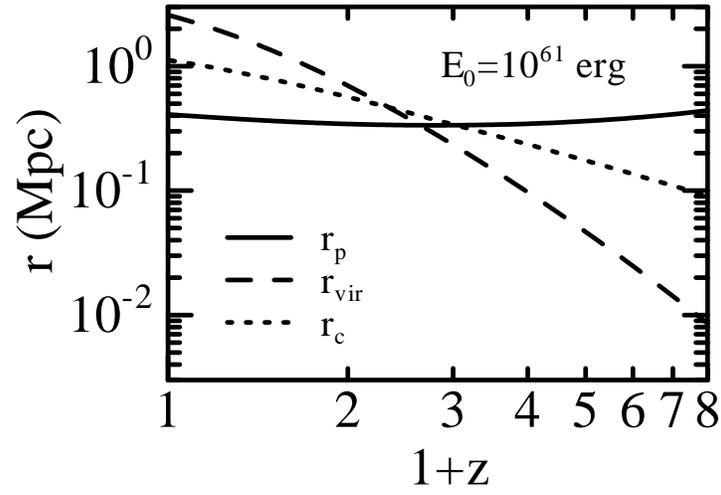}
\caption{The evolutions of the pressure equilibrium radius (a
solid line), virial radius (a dashed line), and cooling radius (a dotted
line) for a cluster with $M_0=10^{15} M_{\sun}$. The explosion energy is
$E_0=10^{61}$~erg. \label{fig:r15}}
\end{figure}

Finally, we make comments on clusters. We have confirmed that the blast
waves driven by quasars are confined in clusters with present mass of
$M_0\gtrsim 10^{15}\; M_{\sun}$ for $z\lesssim 2$
(Figure~\ref{fig:r15}). This may explain the relatively high metal
abundance of X-ray gas in rich clusters \citep{ren97,fuk97}. Moreover,
if quasars are the main heating sources of the X-ray gas in clusters,
shock fronts may be detected at the virial boundary of clusters contrary
to groups. This is because quasars internally heat the X-ray gas at
least for $z\lesssim 2$ and thus the temperature of infalling IGM may be
small \citep[see][]{toz00}.

\acknowledgments
I thank for T. Totani, and S. Inoue for useful comments.

\clearpage

%

\end{document}